\documentclass[runningheads]{llncs}
\usepackage[T1]{fontenc}
\usepackage{cite}
\usepackage{amsmath,amssymb,amsfonts}
\usepackage{graphicx}
\usepackage{textcomp}
\usepackage{booktabs} 
\usepackage{subcaption}
\usepackage{tikz}
\usetikzlibrary{shapes.arrows, positioning}
\usepackage{url}
\usepackage{orcidlink}

\usepackage{listings}
\usepackage{xcolor}

\usepackage{soul}
\setstcolor{red}
 
\usepackage{graphicx} 
\usepackage{pgfplots}
\usepackage{tikz}
\usepackage{cite}

\makeatletter
\newif\if@restonecol
\makeatother

\usepackage[linesnumbered,ruled]{algorithm2e}
\usepackage{algpseudocode}

\usepackage{CJKutf8}
\usepackage{graphicx}
\usepackage[misc]{ifsym}

\begin{document}
\title{A Hybrid Vectorized Merge Sort on ARM NEON}
%
%


\author{Jincheng Zhou\inst{1}~\orcidlink{0009-0007-2281-0767} \and
Jin Zhang\inst{1}~\orcidlink{0000-0002-7464-2247} \and
Xiang Zhang\inst{2,3}~\orcidlink{0000-0002-5201-3802}\and
Tiaojie Xiao\inst{2,3}~\orcidlink{0000-0002-8378-5530}\and
Di Ma\inst{2}~\orcidlink{0009-0000-5794-7906}\and
Chunye Gong\inst{2,3,4(\textrm{\Letter})}~\orcidlink{0009-0002-4825-4686}}

\authorrunning{Jincheng et al.}
%
\institute{School of Computer and Communication Engineering, Changsha University of Science and Technology, Changsha 410114, China \and
College of Computer, National University of Defense Technology, HuNan,Changsha 410073, China
 \and
Laboratory of Digitizing Software for Frontier Equipment, National University of Defense Technology, Changsha 410073,China
\and
National Supercomputer Center in Tianjin, Tianjin 300457, China
\\}

\maketitle              

\vspace{-2mm}

\begin{abstract}
Sorting algorithms are the most extensively researched topics in computer science and serve for numerous practical applications. Although various sorts have been proposed for efficiency, different architectures offer distinct flavors to the implementation of parallel sorting. In this paper, we propose a hybrid vectorized merge sort on ARM NEON, named NEON Merge Sort for short (NEON-MS). In detail, according to the granted register functions, we first identify the optimal register number to avoid the register-to-memory access, due to the write-back of intermediate outcomes. More importantly, following the generic merge sort framework that primarily uses sorting network for column sort and merging networks for three types of vectorized merge, we further improve their structures for high efficiency in an unified \emph{asymmetry} way: 1) it makes the optimal sorting networks with few comparators become possible; 2) hybrid implementation of both serial and vectorized merges incurs the pipeline with merge instructions highly interleaved. Experiments on a single FT2000+ core show that NEON-MS is 3.8 and 2.1 times faster than std::sort and boost::block\_sort, respectively, on average. Additionally, as compared to the parallel version of the latter, NEON-MS gains an average speedup of 1.25.

\keywords{Merge sort  \and Sorting network \and SIMD \and Parallel sort.}
\end{abstract}
\section{Introduction}
Sorting is one of the most extensively studied algorithms in computer science and plays a critical role in various computational applications \cite{1}, such as database retrieval \cite{15}, image processing \cite{28}, and visual computing \cite{29}. Specially, with the advent of the big data era, there is a significantly increasing demand in improving sorting algorithms for data processing.\par

SIMD (Single Instruction, Multiple Data) is one of the most commonly used parallel technologies in modern processors, such as ARM NEON, Intel AVX, and RISC-V Vector Unit. In sorting, SIMD instruction sets are utilized to implement vectorized sorting networks, which handle small-scale data sorting. This is often referred to as in-register sort because all operations are performed on vector registers. Bramas \textit{et al.} \cite{13} explored an efficient quicksort variant on ARM CPUs with Scalable Vector Extensions (SVE-QS). It utilizes the bitonic sorting network for small partitions. Since its small-scale data sorting involves comparing and swapping data (Comparator) within the registers, the interaction across registers directly affects the sorting efficiency. Actually, SVE-QS only uses a few vector registers each time, so the in-register sort runs with the limited SIMD capabilities. Arman \textit{et al.} \cite{11} designed an in-memory merge sort framework, named Origami, which is optimized for scalar operations. Origami stacks all vector registers available to avoid register resource waste, but using too many registers will entail a complex sorting network and could induce unwelcome register-to-memory accesses. Yin \textit{et al.} \cite{12} proposed a highly efficient sorter based on the AVX-512 multi-core architecture. This sorter sorts small-scale data in registers. However, the symmetry merging networks are inefficient.

In terms of the fore-said discussions, on one hand, it is essential to emphasize register-to-memory access rather than register resource utilization. This is because the latter purely uses all the registers, while the former needs to carefully filter out a portion of registers, according to register functionalities. In this way, it is most likely to yield the optimal number of registers for our goal. On the other hand, the simpler the sorting or merging networks, the higher the overall sorting efficiency. Firstly, fewer registers are favorable to simplify the network structure. Secondly, when using the fixed size of registers, few comparators are attractive. Lastly, efficient realization of merging networks could benefit to thread-level parallelism.

\textbullet\, 
We recognize the optimal register number to reduce the register-to-memory access times.

\textbullet\, 
We introduce few-comparator column sort by using the best sorting network \cite{25} with the asymmetric structure, which ensures the optimal sorting network with few comparators to be possible, in contrast to symmetric bitonic or even-odd sorting networks.

\textbullet\, 
We propose a new hybrid bitonic merging network that concurrently implements the serial merge and vectorized merge in an asymmetric fashion, allowing merge instructions to be highly interleaved in the pipeline.

This paper is organized as follows: Section II elucidates our optimization strategy and its implementation. Section III provides an analysis of the results, and Section IV summarizes the paper's key findings and contributions.\par

\section{NEON-MS algorithm}

\subsection{Overview}

\begin{figure}[htbp]
	\centering
	\includegraphics[width=22em]{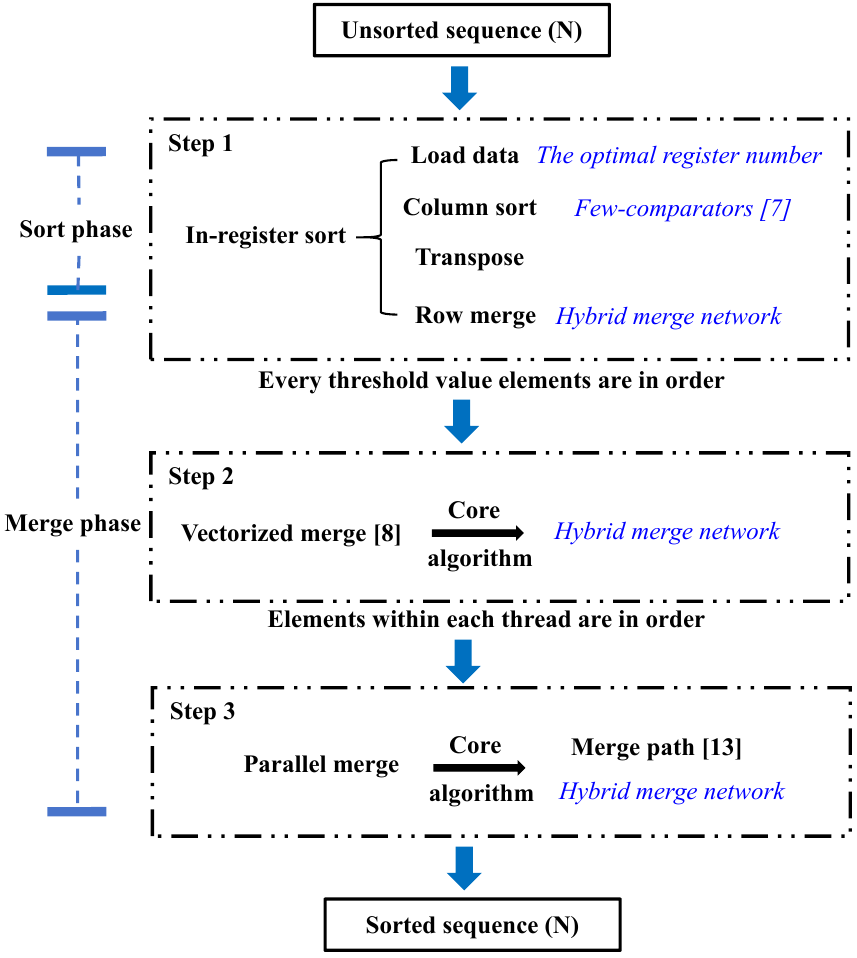}
	\caption{
The flowchart of NEON-MS algorithm, with three improvements highlighted in \textit{\textcolor{blue}{blue italic}}, \emph{i.e.}, the optimal number of the used registers, sorting networks with few comparators for column sort, and three types of merges with our hybrid merging network.
}
	\label{pdrfig1}
    \end{figure}

The overall flowchart of NEON-MS algorithm includes three core components
as shown in Fig. \ref{pdrfig1}. Before detailing them, we are ready to assign the input sequence of the length \textit{N} to all available threads of the size \textit{T}, with each thread being responsible for sorting its allocated subsequence of the length {$\frac{N}{T}$}. A threshold is set to the multiple of the SIMD width. 
When this threshold is achieved, we will employ the improved in-register sort to arrange small subsequences in some order. Then, we propose a hybrid bitonic merging network, using it as the core of vectorized merge \cite{8} to accelerate the merging process. At last, such locally sorted subsequences will be globally merged. We entails a data partitioning strategy \cite{17}. The primary optimization involves balancing the load so that each thread can allocate a comparable amount of workload. In every partitioning block, the thread executes the aforesaid vectorized merge.

\vspace{-5mm}

\vspace{-6mm}

\begin{figure}[htbp]
	\centering
	\includegraphics[width=24em]{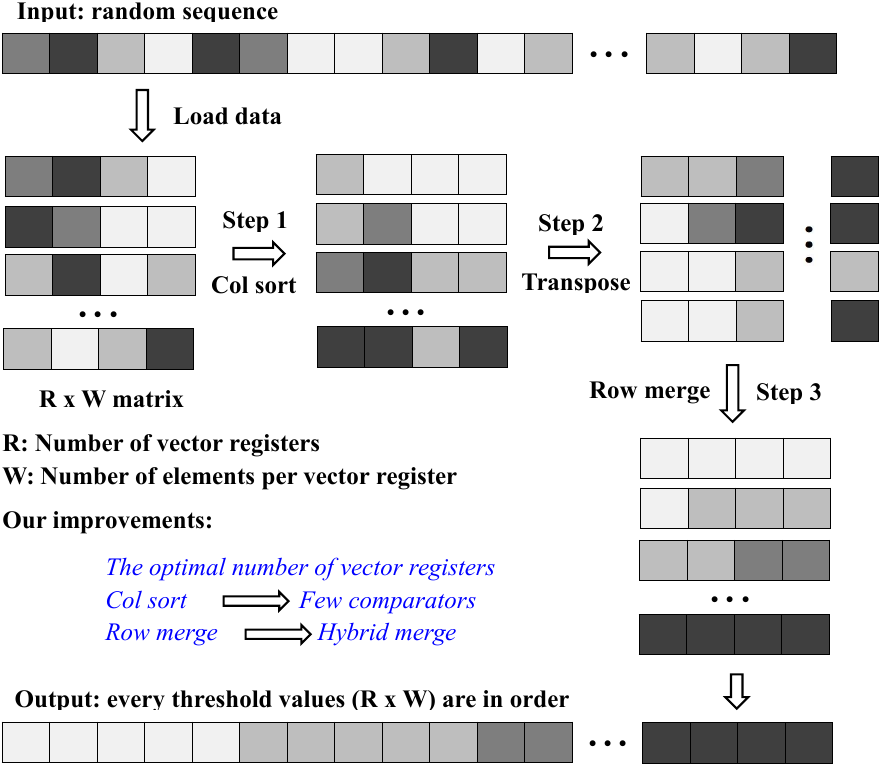}
	\caption{The workflow of the in-register sort ($W$ = 4), where each square represents a data item, with darker cells indicating larger values.}
	\label{pdrfig2}
    \end{figure}

As shown in Fig. \ref{pdrfig2}, the in-register sort consists of four steps: load data, column sort, transpose and row merge. The data launch involves the number of the used registers, and the transpose tunes the register location to make each row sorted. The most important step is column sort, because it not only induces the finest grained sorted subsequences but also offers the transposed yet in-order subsequences to row merge. If the row merge receives the disorder subsequences, the merge outcome is also disorder and meaningless. In light of this insight, it is not surprising that two mergers follows the in-register sort. Both mergers correspond to the vectorized merge and multi-thread parallel merge, respectively. Thus, the row merge of the in-register sort, the vectorized merge and the multi-thread parallel merge share the same merge spirit. In a nutshell, three factors, i.e., the number of the used vector registers, the column sorting network structure, and the merge implementation way, will decide the overall sorting efficiency on ARM NEON.

\vspace{-5mm}

    \vspace{-10mm}

\subsection{The Optimal Register Number}

Since ARM NEON, different from other instruction sets like Intel AVX, gives the shorter vectorized registers, more flexible usage of such registers is a concern. Naturally, the first question is about how many registers used for the in-register sort is suited. Before that, we need to break a long-standing rule that constrain the number of the loaded elements to equal to the square of the number of elements per vector register, i.e., $W\times W$. If we follow this rule in the ARM NEON that has 32 128-bit vector registers, over 85\% of the register hardware resources will be idle and wasted. Recent work \cite{11} addresses this issue by stacking all available $R$ vector registers with the size of $W$. The 32 vector registers in ARM NEON are all used there. 

Once $R$ is set and coupled with the fixed width $W$, the size of the sorting networks is decided. In the subsequent description, we take 32-bit integers as an example, meaning each vector register can store 4 elements, i.e., $W$=4. As one knows, the bigger the $R$, the more complex the sorting networks. For example, the comparators in a 32-element sorting network are three times those in a 16-element one. Besides, the register-to-memory access is always a big computation bottleneck. It is more efficient to store and access data on the registers than to do on the main memory. Thus, the proper value of $R$ not only fits to the simple sorting network structure but also bypasses the register-to-memory access. In ARM NEON, the 16 vector registers meet this need, that is, $R=16$. Our empirical studies in Table \ref{table:x_elements_sorted} show that this case has optimal efficiency on ARM NEON.

\vspace{-4mm}

\subsection{Few-comparators Column Sort}

\vspace{-2mm}

Column sort involves sorting the same channel across multiple vector registers, which is a key utilization of vectorized properties. Most implementations use the sorting networks due to their simple welding structure. 
\emph{Which sorting network structure for column sort are most efficient?} 
The efficiency can be evaluated by the number of the comparators used in a sorting network. Table \ref{tablettt} lists the relation between the number of the comparators and sorting networks of varying input size $n$. When $n=2^i$, wherein $i$ is a positive integer, the bitonic and odd-even sorting networks have the symmetric structures and the fixed comparators. Most current implementations belong to this sort but do have no room to further optimize the number of the comparators. In contrast, the asymmetric structure has fewer comparators than the symmetric structure, especially when the input size exceeds 8, because the upper bound of the comparators' number is less than its symmetric siblings. Besides, the odd input sizes are incompatible with the ready-to-use transpose operation, so the respective sorting networks are rarely used by the in-register sort. For larger networks, there exist the proven optimal lowest bounds for the number of comparators \cite{26}. In ARM NEON, larger sorting networks ($n>$32) is infeasible, because the total register number available is 32 at most.

\vspace{-7mm}

\begin{table}[h]
\centering
\caption{Number of comparators in different sorting networks of input size $n$.}
\label{tablettt}
\vspace{2mm}
\begin{tabular}{|c|c|c|c|}
\hline
$n$ & Bitonic & Odd-even & Asymmetric Network \\ \hline
4 & 6 & 5 & 5 \\ \hline
8 & 24 & 19 & 19 \\ \hline
16 & 80 & 63 & $55\sim60$ \\ \hline
32 & 240 & 191 & $135\sim185$ \\ \hline
\end{tabular}
\end{table}

\vspace{-6mm}

Since the optimal value of $R$ is 16, namely, $n=16$, according to TABLE 1, we have the chance to reduce the number of the comparators. Thus, we introduce the 16-element best sorting network \cite{25} for efficient multi-column sorting.
This network has fewer comparators than the commonly-used symmetric sorting networks. This saves time.

Through our few-comparators column sort, the locally sorted sequences are distributed by column across different registers. This data layout being order in column must be recovered into the sorted layout in a register (in row) via a matrix transpose before the data are written back to memory. If $W$$<$$R$, we regard the $W\times W$ matrix transpose as the base matrix transpose like the atomic operation. Then transposing an asymmetric matrix $R\times W$ can be reduced into multiple small base matrix transpose. This involves adjusting the positions of the vector registers, with few overheads.

\subsection{Hybrid Bitonic Merger}

\subsubsection{Motivation.}
After column sort and transpose, each ordered subsequences need to be further processed by multiple mergers to make the full sequence in order. As previously mentioned, three mergers including in-register row merge, vectorized merge and parallel merge share the same merge spirit. In ARM NEON, a proper candidate for the merge is the bitonic merging network, due to its simplicity and efficiency. Nevertheless, there are two existing ways to implement bitonic merging network. The first way straightly follows the predefined merging network (See Fig. \ref{32pdrfig}). Due to the presence of the compare and swap operations in the comparator, the underlying assembly codes contain some branch jump instructions (See Fig. \ref{fig:combined1}). As one knows, once the branch jump prediction meets the errors, this leads to extra execution cycles. Seemingly, these branch jump instructions can be transformed into conditional swap instructions by using ternary operations (See Fig. \ref{fig:combined2}), but each time only one conditional instruction is allowed. In contrast, the second vectorization implementation can concurrently execute multiple comparators for efficient merging. Before each comparison operation, the data needs to be shuffled so that correct comparisons can be obtained. However, the shuffle operation between vector registers in ARM NEON is not sufficiently flexible because it requires additional type conversion operations, which incurs more instruction overheads. Thus, both of implementations have individual pros and cons. This hints that we could enjoy their strengths.\par

\vspace{-6mm}

\begin{figure}[htbp]
    \centering

    \begin{subfigure}{\textwidth}
        \centering
        \caption{Branch jump instruction (\textbf{b.le})}
        \label{fig:combined1}
        \begin{minipage}{.5\textwidth}
            \centering
            \begin{algorithm}[H]
                \small
                \SetKwProg{Fn}{Function}{:}{end}
                \Fn{$Comparator\_v_0$($a,l,r$)}{
                    $if$ $(a[l] > a[r])$\\
                    $\ $ $std::swap(a[l], a[r])$\;
                }
            \end{algorithm}
        \end{minipage}%
        \hfill
        \begin{minipage}{.5\textwidth}
            \lstset{basicstyle=\ttfamily\small}
            \begin{lstlisting}
            ...
28a8:6b04007f cmp  w3,w4
28ac:5400006d b.le 28b8 
            ...
            \end{lstlisting}
        \end{minipage}
    \end{subfigure}

    \begin{subfigure}{\textwidth}
        \centering
        \caption{Conditional swap instruction (\textbf{csel})}
        \label{fig:combined2}
        \begin{minipage}{.5\textwidth}
            \centering
            \begin{algorithm}[H]
                \small
                \SetKwProg{Fn}{Function}{:}{end}
                \Fn{$Comparator\_v_1$($a,l,r$)}{
                    $bool$ $flag = (a[l] > a[r])$\;
                    $int$ $temp = a[l]$\;
                    $a[l] =$ $r?$ $a[r] : a[l]$\;
                    $a[r] =$ $r?$ $temp : a[r]$\;
                }
            \end{algorithm}
        \end{minipage}%
        \hfill
        \begin{minipage}{.5\textwidth}
            \lstset{basicstyle=\ttfamily\small}
            \begin{lstlisting}
            ...
2878:6b03009f cmp  w4,w3
287c:1a83d084 csel w4,w4,w3,le
            ...
            \end{lstlisting}
        \end{minipage}
    \end{subfigure}

    \caption{Two implementations of a comparator: the source code on the left, and the respective core assembly code on the right.}
    \label{fig:combined}
\end{figure}

\vspace{-11mm}

\subsubsection{Hybrid Implementation.} 
\begin{figure}[!htbp]
    \centering
    \includegraphics[width=0.7\textwidth]{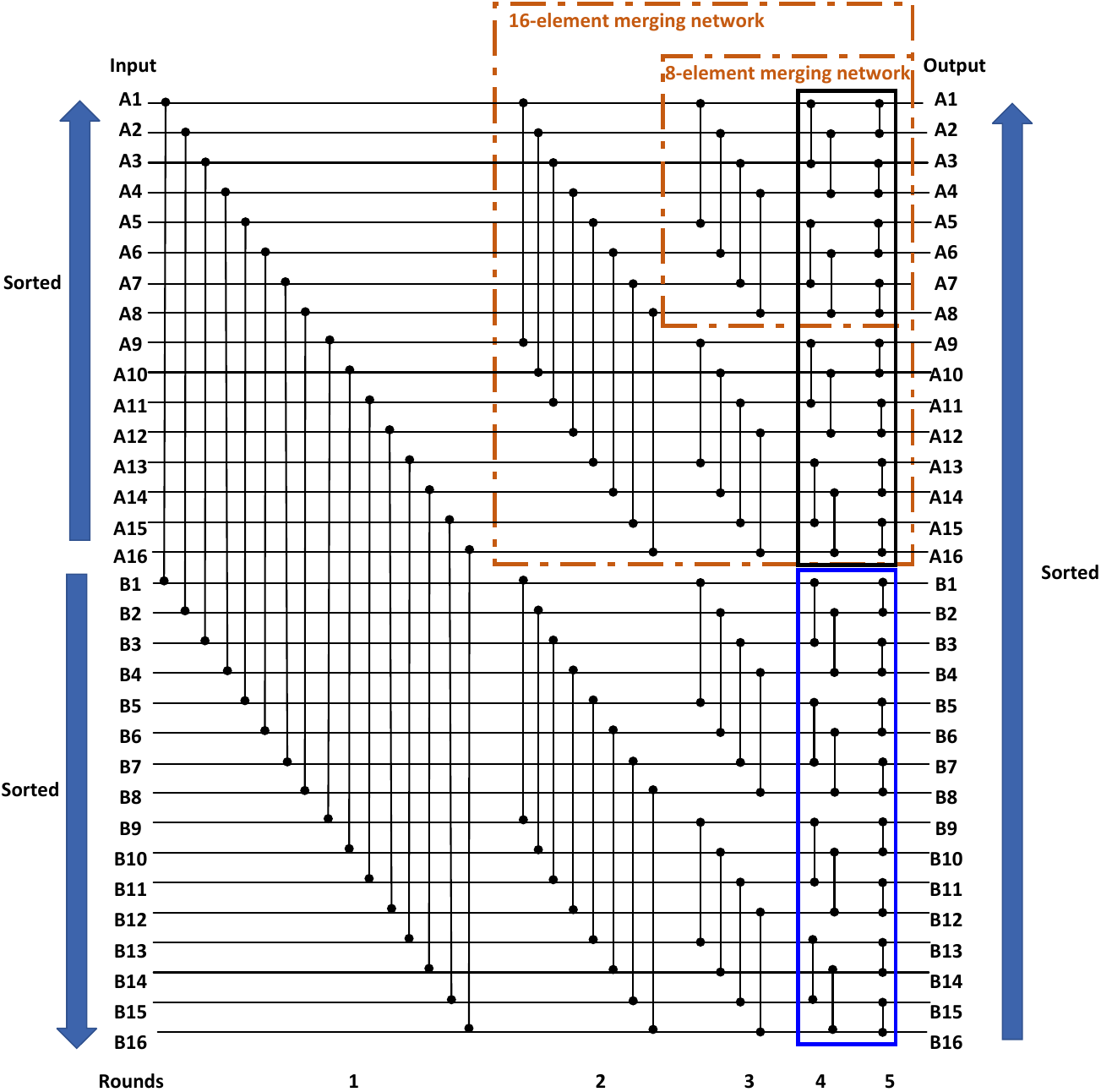}
    \caption{A 32-element bitonic merging network. The \textbf{black} and \textbf{{\color{blue}blue}} rectangles represent data swapping in vector registers, characterized by their symmetry and independent operation.}
    \label{32pdrfig}
    \end{figure}

Recall that the bitonic merging network in itself has symmetric structures, as Fig. \ref{32pdrfig} shows. Then, a 32-element bitonic merging network has two 16-element symmetric parts, while each 16-element bitonic merging network has two 8-element symmetric parts, etc. 
Obviously, if necessary, such symmetrical parts can be implemented in any one of two previous ways and then run in parallel. By this insight, we propose a hybrid bitonic merging network that concurrently utilizes the above two implementation ways in the tail of the merging network, as the \textbf{black} and \textbf{{\color{blue}blue}} rectangles of Fig. \ref{32pdrfig} show. Such a hybrid implementation, with the help of the compiler, enable the assembly instructions of both the serial and the vectorized implementations to interleave with each other in the pipeline. This benefits to reducing waiting time on the conditional instructions and the times of data swapping among vector registers. \par

This hybrid spirit will serve for our three mergers which are used in the in-register sort, vectorized merge and multi-thread parallel merge, repsectively.

\vspace{-3mm}

\vspace{-5mm}

\vspace{-4mm}

\section{Experimental results}

In this section, we evaluate NEON-MS on the FT2000+ processor, which operates at 2.3 GHz and features 64 cores. The processor architecture includes a shared 2M L2 cache for every group of 4 cores. Additionally, every 8 cores form a NUMA node, which is managed by DDR4 memory control. NEON-MS is implemented using the C++ language, taking advantage of ARM NEON features for optimization. In the parallel version, we employ the OpenMP standard. All implementations are compiled using GCC 9.3.0 with \textit{-O3} level optimization.\par

First, we analyze the optimization effects of NEON-MS at different stages. As sorting small-scale data is very fast, we calculate average performance through multiple iterations. In our tests of in-register sort, we traversed 64K random integers and iterated 100 times. Second, we compare the performance of NEON-MS with that of boost::block\_sort—one of the most efficient functions in C++ sorting implements \cite{22}, std::sort—a widely used sorting function in the C++ standard library in a single-thread context. Finally, we test the parallel performance of NEON-MS and boost::block\_sort.

\vspace{-2mm}

\subsection{Localized performance}

\vspace{-10mm}

\begin{table}[h!]
\centering
\renewcommand{\arraystretch}{1.1} 
\caption{\textcolor{black}{
The running time ($\mu s$) for sorting $X$ elements in an $R$$\times$ 4 matrix.
}
}
 \vspace{1em}

\begin{tabular}{|c|c|c|c|c|c|}
\cline{2-6}
\multicolumn{1}{c|}{} & \multicolumn{5}{c|}{Every $X$ elements are in order} \\ \hline
$R$ & $X$=4 & $X$=8 & $X$=16 & $X$=32 & $X$=64 \\ \hline
4 & 38 & 105 & 186 & - & - \\ \hline
8 & - & 49 & 112 & 179 & - \\ \hline
16 & - & - & 76  & 134  & 203  \\ \hline
\textbf{16*} & - & - & \textbf{65} & \textbf{121} & \textbf{183} \\ \hline
32 & - & - & -  & 128  & 194  \\ \hline
\end{tabular}
\label{table:x_elements_sorted}
\smallskip 
\par \noindent\footnotesize  \textbf{16*} indicates the best 16-element sorting network.
\end{table}

\vspace{-5mm}
\textcolor{black}{
Table \ref{table:x_elements_sorted} displays the running times required for sorting every $X$ elements in in-register sort across various numbers of vector registers.
}It is evident that using 16* vector registers yields the best running time, primarily because this setup utilizes a best network designed for 16 elements, which requires fewer comparisons. Moreover, it is observable that as the number of vector registers increases, the time required to sort the same quantity of $X$ sorted elements decreases (excluding the 16*). This improvement occurs because more data can be processed simultaneously within each in-register sort. Although $R$ = 32 demonstrates better performance than $R$ = 16, due to its complexity, we do not consider it. Therefore, we conclude that the optimal number of vector registers is 16*.\par

\vspace{-8mm}


\begin{table}[h!]
\centering
\renewcommand{\arraystretch}{1.1} 
\caption{The relationship between merge lengths and merging speeds (elements/$\mu s$) across two merging methods.}
\label{tablewww}
 \vspace{1em}
\label{table:sort_comparison}
\begin{tabular}{|c|c|c|c|}
\hline
Merge Length $\rightarrow$ & 2x8 $\rightarrow$ 16 & 2x16 $\rightarrow$ 32 & 2x32 $\rightarrow$ 64 \\ \hline
Vectorized Bitonic & 873.81 & 1024 & \textbf{897.75} \\ \hline
Hybrid Bitonic & \textbf{1057.03} & \textbf{1092.27} & 840.21 \\ \hline
\end{tabular}
\end{table}

\vspace{-4mm}

Table \ref{tablewww} shows the merging speeds for different merge lengths across two merging methods. It can be observed that the merge speed of the hybrid bitonic method is faster than that of the vectorized bitonic method for merge lengths of 8 and 16. This advantage arises because our hybrid bitonic merging network concurrently utilizes two implementation ways in the tail of the merging network. This benefits to reducing waiting time on the conditional instructions and the times of data swapping among vector registers. For larger merge lengths, however, the merge speed of hybrid bitonic is slower than that of vectorized bitonic. This performance decrease is due to the fact that serial implementation generates temporary data that cannot be stored in the limited registers available. These data must be stored in memory, resulting in increased overhead.\par

\subsection{Overall performance}

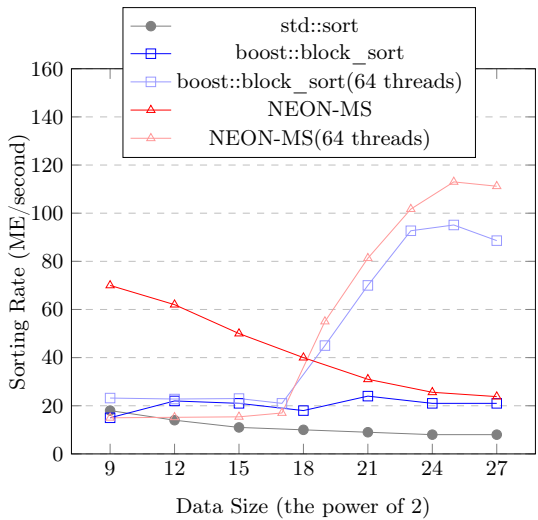
\begin{figure}[htbp] 
\centering 
\begin{tikzpicture}[scale=0.9]
\begin{axis}[
    xlabel={Data Size (the power of 2)},
    ylabel={Sorting Rate (ME/second)},
    ylabel near ticks,
    ylabel shift=-6pt,
    ymin=0, 
    ymax=160, 
    ytick={0,20,40,60,80,100,120,140,160},
    yticklabels={0,20,40,60,80,100,120,140,160},
    xtick={9,12,15,18,21,24,27},
    legend pos=north east,
     legend style={font=\footnotesize, at={(0.11,1.16)}, anchor=north west},
    ymajorgrids=true,
    grid style=dashed,
]

\addplot[
    color=gray,
    mark=*,
] coordinates {
    (9,18)
    (12,14)
    (15,11)
    (18,10)
    (21,9)
    (24,8)
    (27,8)
};
\addlegendentry{std::sort}

\addplot[
    color=blue,
    mark=square,
] coordinates {
    (9,15)
    (12,22)
    (15,21)
    (18,18)
    (21,24)
    (24,21)
    (27,21)
};
\addlegendentry{boost::block\_sort}

\addplot[
    color=blue!40,
    mark=square,
] coordinates {
    (9,23.2)
    (12,22.8)
    (15,23)
    (17,21)
    (19,45)
    (21,70)
    (23,92.7)
    (25,95.1)
    (27,88.6)
};
\addlegendentry{boost::block\_sort(64 threads)}

\addplot[
    color=red,
    mark=triangle,
] coordinates {
    (9,70)
    (12,62)
    (15,50)
    (18,40)
    (21,31)
    (24,25.6)
    (27,23.8)
};
\addlegendentry{NEON-MS}

\addplot[
    color=red!40,
    mark=triangle,
] coordinates {
    (9,15)
    (12,15.2)
    (15,15.4)
    (17,17)
    (19,55)
    (21,81.3)
    (23,101.7)
    (25,113)
    (27,111.2)
};
\addlegendentry{NEON-MS(64 threads)}

\end{axis}
\end{tikzpicture}
\caption{Sorting Rate (ME/s: million elements per second) of different sorting methods for different data sizes. } 
\label{figggg}
\end{figure}

Fig. \ref{figggg} shows the performance of various sorting algorithms from 512K to 128M data sizes. This figure indicates that the overall performance of NEON-MS is better than other two methods. It achieves performance ranging from 23.5 to 70 ME/s, averaging 2.1 times faster than boost::block\_sort, and 3.8 times faster than std::sort. In addition, NEON-MS is 1.25  times faster than boost::block\_sort with 64 threads for large-scale data. This is attributed to the advantages of our parallel merge strategy, where each available thread remains active. For small data scales, the performance of NEON-MS with 64 threads is poorer than that of the boost::block\_sort. This is because the creation of parallel workloads, thread synchronization, and extra computing tasks, such as the calculation of segmentation points in our implementation, occupy a major portion of the execution time. Meanwhile, boost::block\_sort, while also a merging algorithm with multi-thread environments, features a small auxiliary memory (block\_size multiplied by the number of threads). This configuration substantially minimizes the consumption of additional memory, leading to enhanced sorting efficiency.\par

\section{Conclusion}

This paper proposes a hybrid vectorized merge sort on ARM NEON architecture. This implementation utilizes the structural features of FT2000+ processors. 
In the sort phase, we use the optimal number of registers to achieve a simplified, effiecient sort and introduce the best sorting network to futher enhance sorting efficiency. During the overall merge phase, we propose a new hybrid bitonic merging network that allows merge instructions to be highly interleaved in the pipeline.
The results show that single-thread NEON-MS performs, on average, 3.8 and 2.1 times faster than std::sort and boost::block\_sort, respectively. Additionally, the multi-thread NEON-MS demonstrates a good performance enhancement.\par

\section{Acknowledgements}
This research was supported by the National Natural Science Foundation of China (Grant Nos. 62032023, 42104078 and 6190241).

%
%
%

\bibliographystyle{plain}

\bibliography{reference}





\end{document}